\documentclass[%
 reprint,
nofootinbib,
 amsmath,amssymb,
 aps,
floatfix,
]{revtex4-1}

\usepackage{xspace}
\usepackage{xcolor}
\usepackage{caption}
\usepackage[normalem]{ulem}
\usepackage[breaklinks, plainpages=false, colorlinks=true, anchorcolor=cyan, linkcolor=magenta, citecolor=cyan, urlcolor=purple, bookmarks=false]{hyperref}
\usepackage{appendix}
\usepackage{graphicx}
\usepackage{acronym}
\usepackage{bm}
\usepackage{subcaption}
\usepackage{bbold}
\usepackage{tabularray}
\usepackage{soul}

\newacro{imbh}[IMBH]{intermediate-mass black hole}
\newacro{smbh}[SMBH]{supermassive black hole}
\newacro{bhns}[BHNS]{black hole neutron star}
\newacro{bbh}[BBH]{binary black hole}
\newacro{bh}[BH]{black hole}
\newacro{bns}[BNS]{binary neutron star}
\acrodef{FAR}[FAR]{false alarm rate}
\newacro{bf}[BF]{Bayes' factor}
\newacro{cbc}[CBC]{compact binary coalescence}
\newacro{ce}[CE]{Cosmic Explorer}
\acrodef{SNe}[SNe]{Supernova}
\newacro{da}[DA]{data analysis}
\newacro{et}[ET]{Einstein Telescope}
\newacro{eob}[EOB]{Effective-One-Body}
\newacro{fd}[FD]{frequency domain}
\newacro{gw}[GW]{gravitational wave}
\newacro{gr}[GR]{General relativity}
\newacro{hm}[HM]{Higher mode}
\newacro{ifo}[IFO]{interferometer}
\newacro{imr}[IMR]{inspiral-merger-ringdown}
\newacro{im}[IM]{inspiral-to-merger}
\newacro{kagra}[KAGRA]{Kamioka Gravitational Wave Detector}
\newacro{ligo}[LIGO]{Laser Interferometer Gravitational-Wave Observatory}
\newacro{lso}[LSO]{Last Stable Orbit}
\newacro{lvc}[LVC]{LIGO-Virgo Collaboration}
\newacro{lvk}[LVK]{LIGO-Virgo-Kagra Collaboration}
\newacro{lo}[LO]{leading order}
\newacro{ns}[NS]{neutron star}
\newacro{nr}[NR]{numerical relativity}
\newacro{pn}[PN]{post-Newtonian}
\newacro{pe}[PE]{parameter estimation}
\newacro{psd}[PSD]{power spectral density}
\newacro{cwb}[cWB]{coherent waveburst}
\newacro{far}[FAR]{false alarm rate}
\newacro{ifar}[iFAR]{inverse false alarm rate}
\newacro{ml}[ML]{machine learning}
\newacro{cnn}[CNN]{convolutional neural network}
\newacro{asd}[ASD]{amplitude spectral density}
\acrodef{KN}[KN]{kilonova}
\newacro{xg}[XG]{next-generation}
\newacro{jsd}[JSD]{jensen shannon divergence}
\newacro{qnm}[QNM]{quasi-normal mode}
\newacro{cwt}[CWT]{continuous wavelet transform}
\newacro{hlv}[HLV]{Hanford-Livingston-Virgo}

\acrodefplural{KN}[KNe]{kilonovae}
\newacro{qc}[QC]{quasi-circular}
\newacro{snr}[SNR]{signal-to-noise ratio}
\acrodef{SNR}[SNR]{signal-to-noise ratio}
\newacro{ng}[NG]{Next Generation}
\newacro{eos}[EoS]{Equation of State}

\newcommand{\jf}[1]{{\textcolor{magenta}{{JF: #1}} }}
\newcommand{\kc}[1]{{\textcolor{blue}{{KC: #1}} }}

\begin{document}

\title{Improving the detection significance of gravitational wave transient searches with CNN models} 

\author{Johann Fernandes}
    \email{johann.fernandes@iitb.ac.in}
    \affiliation{Department of Physics, Indian Institute of Technology, Bombay, Powai, 400076, India}
\author{Archana Pai}%
    \email{archanap@iitb.ac.in}
    \affiliation{Department of Physics, Indian Institute of Technology, Bombay, Powai, 400076, India}
\author{Koustav Chandra}
    \email{kbc5795@psu.edu}
\affiliation{Institute for Gravitation and the Cosmos, Department of Physics and Department of Astronomy and Astrophysics, The Pennsylvania State University, University Park, PA 16802 USA}%

\date{\today}

\begin{abstract}
    Gravitational wave (GW) transient searches rely on signal-noise discriminators to distinguish astrophysical signals from noise artefacts. These discriminators are typically tuned towards expected signal morphologies, which may limit their effectiveness as detector sensitivity improves and more complex signals, such as from core collapse supernovae or compact binary mergers featuring precession, higher-order harmonics, or eccentricity, become detectable. In this work, we use a Convolutional Neural Network-based approach to classify noise transients from astrophysical transients, aiming to enhance the sensitivity of existing searches. We evaluate our method on two matched filter based searches, PyCBC-IMBH and PyCBC-HM tuned for \ac{imbh} binary systems. Our approach improves the sensitive volume-time reach of these searches by approximately 30\% at a false alarm rate of once per 100 years. Finally, we apply our method to the first four chunks of the first half of the third observation run and demonstrate a marked improvement in significance. In particular, we significantly improve the first IMBH binary \ac{gw} event GW190521 with an IFAR exceeding 42000 yr.
\end{abstract}

\maketitle

\section{Introduction}


The \ac{lvk} has observed approximately 100 \ac{gw} transient events in the first three observational runs, with most of them being consistent with merging quasi-spherical compact objects~\citep{LIGOScientific:2014pky, VIRGO:2014yos, KAGRA:2018plz, LIGOScientific:2018mvr, LIGOScientific:2020ibl, KAGRA:2021vkt}. This number is projected to at least double during the ongoing fourth observing run (O4) of the advanced \ac{lvk} detectors and increase by orders of magnitude with the advent of \ac{xg} \ac{gw} detectors, such as \ac{ce} and \ac{et}~\citep{Punturo:2010zz, Reitze:2019iox, Maggiore:2019uih, Evans:2021gyd, Branchesi:2023mws, Gupta:2023lga, KAGRA:2013rdx, Hild:2010id, Abernathy:2011, LIGOScientific:2016wof}. 

These detections, primarily \ac{bbh} mergers, have provided a unique opportunity to study \acp{bh}, enabling precision tests of \ac{gr} in the high-curvature, strong-field regime and offering insights into the astrophysical origins and nature of \ac{bh} populations \citep{LIGOScientific:2021sio, KAGRA:2021duu, LIGOScientific:2021usb}.

However, detecting and inferring the properties of these events becomes challenging, especially when the signal is short-lived. These short-lived signals can be either intrinsic, like supernova explosions, or a result of the limited bandwidth of the detector. In both cases, detecting such signals poses significant challenges due to the noise transients inherent in interferometric detectors. These non-Gaussian noise transients, also called glitches, are of terrestrial origin \cite{LIGO:2021ppb} and are classified based on their morphology and/or source. Common categories include blips, scattered light, koi fish, tomtes, etc. \citep{Cabero:2019orq, Zevin:2016qwy}. Mitigating these noise transients and improving the detection sensitivity of searches is an active area of research.

As an example, consider the case of \ac{cbc} systems with heavier masses, especially when one of the component masses or the final \ac{bh} is an \ac{imbh}. The \ac{gw} signal from such a system falls within the lower frequency range of the advanced \ac{lvk} detectors. This results in a short-lived \ac{gw} transient predominantly characterised by the merger and ringdown phases, with only a tiny contribution from the inspiral phase. 
Most notably, the first confident \ac{imbh} binary event GW190521 ~\citep{LIGOScientific:2020iuh} observed during the third observation run of \ac{lvk} detectors was of this type, being barely observable within the detector bandwidth. As a result, GW190521 has been the subject of multiple interpretations, the most credible being a quasi-spherical \ac{imbh} binary event. However, other analyses have also suggested possibilities that range from a hyperbolic encounter to the result of a head-on collision between exotic compact objects ~\citep{Fishbach:2020qag, Nitz:2020mga, Romero-Shaw:2020thy, Gayathri:2020coq, CalderonBustillo:2020xms, Estelles:2021jnz, Olsen:2021qin, CalderonBustillo:2022cja, Chandra:2023nge}.


Consequently, template-based searches like PyCBC-broad, which use quasi-circular quadrupolar \ac{bbh} waveforms, failed to identify this event with sufficient significance (see supplement of \citep{LIGOScientific:2020iuh}). 
Additionally, the PyCBC-broad search suffered from the ``look-elsewhere effect", as it spanned a broad parameter space of \ac{cbc} systems ~\citep{Chandra:2020ccy}. As a result, its empirically derived background included noise triggers resembling GW190521. To overcome this challenge, the PyCBC-IMBH search was developed. It was specifically designed to target quasi-circular \ac{bbh} signals with a detector-frame total mass $M_T(1+z)/M\odot>100$ \citep{Chandra:2021wbw, Chandra:2021xvs}, incorporating strict vetoes to suppress glitches. Meanwhile, the PyCBC-HM search \citep{Chandra:2022ixv} was tailored for mass-asymmetric edge-on intermediate-mass binary black hole (IMBH) systems. PyCBC-HM was the first matched-filter based approach for asymmetric IMBH binaries by incorporating higher-order modes into its quasi-circular template bank.

To summarise, template-based searches often devise different signal-noise discriminators and incorporate them in the test statistics to downrank the noise triggers and improve the search's signal-noise discriminatory ability \citep{Allen:2004gu, Nitz:2017lco, Messick:2016aqy} 
This requires dedicated efforts to identify new signal-noise discriminators that are independent of the existing ones.

Weakly modelled searches, such as \ac{cwb}, on the other hand, utilise 
 the excess coherent power across a network of detectors, to detect signals. They make minimal assumptions about the signal morphology in the time-frequency plane. As with templated searches, they use a variety of signal-noise discriminators during the post-processing stage to compute the background and thus, the final significance of the \ac{gw} triggers \citep{Klimenko:2015ypf}.

As detector sensitivity improves, we anticipate observing more \ac{gw} events that are currently very rare or remain unobserved \citep{Hall:2022dik, ET:2019dnz}. However, we may also see additional glitches that could potentially display complex morphologies due to the larger frequency bandwidth of these detectors. Without effective signal-noise discriminators, we risk missing these exceptional events amidst the noise.
 It is crucial to improve searches as rare events can provide insights into the astrophysical properties of sources. For instance, \ac{imbh} binaries can probe black hole spectroscopy via the remnant \citep{Ghosh:2021mrv, Ota:2021ypb}, whereas kick velocities can inform us about the astrophysical formation channel of massive \ac{bh}s \citep{Mahapatra:2021hme, Fishbach:2017dwv, Gerosa:2017kvu}. These can have a direct implication on astrophysics and \ac{bh} physics.

Over the past decade, several \ac{ml}-based \ac{gw} algorithms, particularly those based on deep learning, have proven effective in rapid signal detection, parameter estimation, and glitch classification, among many other applications. For instance, \citep{Marx:2024wjt} utilise a modified ResNet architecture to detect signals in timeseries data. Rapid inference of signal parameters has also been demonstrated through neural posterior estimation, offering speedups of about $10^3-10^4$ times over traditional Bayesian methods while maintaining agreement with them \citep{Green:2020hst, Green:2020dnx, Dax:2021tsq}. In addition, methods such as gradient boosting and Gaussian Mixture Modelling have also been used as postprocessing steps for \ac{cwb} leading to an improved detection efficiency \citep{Szczepanczyk:2022urr, Smith:2024bsn}. Generative models such as Gengli have been developed to create realistic glitches, helping to create synthetic datasets that more closely resemble real detector data \citep{Lopez:2022lkd}. A broader overview of ML applications in GW astronomy can be found in \citep{Cuoco:2024cdk}.

In this article, we use a \ac{cnn} based signal-noise discriminator to improve existing \ac{gw} transient searches by modifying their detection statistic.
We use a fine-tuned Inception-V3 network output to augment the ranking statistic. To demonstrate the performance, we use the optimised PyCBC based IMBH searches, which fall in the category of the short duration \ac{gw} transient searches and show that the existing optimised searches have scope to improve after incorporating the signal-noise discriminator. We show that, on average, the volume reach is improved by $\sim 35 \%$ compared to existing optimised PyCBC searches. Though here, we use specific matched filter searches, the method is general enough to be adapted for any short-duration \ac{gw} transient search.

The paper is organised as follows. In sec \ref{motives}, we introduce the methodology of our approach with sec \ref{cwt} and \ref{methods} touching upon the generation of the time-frequency maps and the model architecture, respectively. In sec \ref{ranking} we describe the metric obtained from the \ac{cnn} model outputs that will improve the detection statistics. Finally, in sec \ref{application}, we will apply the model to two PyCBC based searches, discussing the improvements in sensitivity obtained.

\section{Methodology} \label{motives}



Glitches predominantly impact \ac{gw} searches targeting burst signals—transients with limited duration and frequency bandwidth. To address this, we exploit the distinct morphological differences between glitches and astrophysical bursts in the time-frequency plane and thus perform our analysis in the time-frequency domain. In particular, we use \ac{cnn}'s ability to identify and differentiate images to distinguish noise and signal triggers.


We first identify a suitable search to augment with our model and then apply our \ac{cnn} model on time-frequency maps of triggers obtained from this search. These \ac{cnn} outputs are then used to reweigh the detection statistic of the base search, a metric which is used to compute the significance of any signal detection algorithm. This integration aims to improve the significance of incoming \ac{gw} signals buried within the detector noise, potentially enhancing the detection of weak signals. For our \ac{cnn} model, we chose to fine-tune an Inception-V3 network on our training dataset \citep{szegedy2016rethinking}. This allows for faster model convergence since the model is initialised with pre-trained weights obtained after training on the ImageNet dataset.

   \subsection{Continuous Wavelet Transforms} \label{cwt}

\begin{figure}[htb]
   \makebox[\linewidth][c]{%
   \centering
\begin{subfigure}{0.4\textwidth}
   \includegraphics[width=0.5\textwidth]{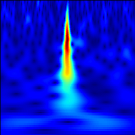}
   \caption{Blip}
   \label{blip}
\end{subfigure}
\hspace{-3.5cm}
\begin{subfigure}{0.4\textwidth}
   \includegraphics[width=0.5\textwidth]{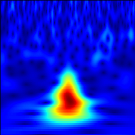}
   \caption{Tomte}
   \label{tomte}
\end{subfigure}
}
\makebox[\linewidth][c]{%
\begin{subfigure}{0.4\textwidth}
   \includegraphics[width=0.5\textwidth]{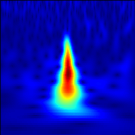}
   \caption{Low Frequency Blip}
   \label{blip_low_freq}
\end{subfigure}
\hspace{-3.5cm}
\begin{subfigure}{0.4\textwidth}
   \includegraphics[width=0.5\textwidth]{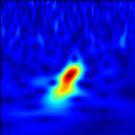}
   \caption{IMBH binary}
   \label{imbh}
\end{subfigure}
}
\caption{Example CWT maps used for training.}
\label{fig:cwt_maps}
\end{figure}

    We obtain \ac{cwt} maps using the analytic Gabor-Morlet wavelet $W_\pi$ defined as follows:
    \begin{equation}
        W_\pi\left(f_0, \eta\right)=\frac{1}{\pi^{1 / 4}}\left(e^{2 i \pi f_0 \eta}-e^{-\left(2 \pi f_0\right)^2 / 2}\right) e^{-\eta^2 / 2}
    \end{equation}
    where $f_0$ is the central frequency and $\eta = (t-\tau)/a$ with $\tau$ and $a$ being the time-translation and scaling parameter~\citep{Flandrin:2018}. Given the short duration of glitches and short duration \ac{gw} transients, we generate \ac{cwt} maps within a 200 ms window around the trigger time. The lower frequency cutoff is set to 15 Hz, as lower frequencies are dominated by detector noise. We set an upper-frequency cutoff of 512 Hz. Fig \ref{fig:cwt_maps} shows some examples of \ac{cwt} maps fed into the network. The glitches \ref{blip}, \ref{tomte} and \ref{blip_low_freq} show clear burst like behaviour that can mimic \ac{gw} transients (\ref{imbh}), contaminating search pipelines. However, the time frequency maps also show subtle differences in morphology, such as the spreads in either the time or frequency axis that a \ac{cnn} can use to distinguish between them.

   
   \subsection{Model architecture and training methodology} \label{methods}
  This section summarises our \ac{cnn} model. We use the Inception-V3 network to classify CWT maps. The first step is curating our training dataset towards the \ac{gw} transients we are targeting. In particular, we need to choose glitches primarily triggered by the base search algorithm. This is done by a visual inspection of the CWT maps of the background triggers, where we identify prominent glitch classes. We then use Gravity Spy to obtain instances of these glitches in real LIGO data \citep{Zevin:2016qwy}. Gravity Spy is a citizen science project that uses human volunteers to classify and identify new glitch classes from a collection of noise triggers. These human-vetted samples are used to train a \ac{cnn} as a classifier. Each glitch classified by Gravity Spy is assigned a confidence score. To ensure that glitches from other classes do not contaminate our training dataset due to Gravity Spy misclassifications, we restrict ourselves to choosing images that have a confidence exceeding 90\%. In addition to these glitch classes, we add another class with \ac{cwt} maps of pure Gaussian noise coloured with the \ac{psd} obtained from the third observing run (O3) of the Hanford and Livingston detectors. Finally, for the signal class, we construct a custom training dataset based on the target parameter space, which we add to the coloured Gaussian noise.\\

  We use this curated training dataset to fine-tune the Inception-V3 model for our use case. The network thus consists of the Inception-V3 model followed by a dense layer with a 30\% dropout to avoid overfitting. These are connected to a final layer with the number of neurons determined by the number of classes being considered. The final layer is normalised with a softmax function. In general, for a layer with $n$ units with outputs denoted by $x_1, x_2, ...,x_n$ the softmax function is given by.
   \begin{align*}
       \boldsymbol{\sigma}:\mathbb{R}^n &\longrightarrow (0,1)^n\\
       (x_1,x_2, ..., x_n) &\longmapsto \boldsymbol{\sigma}(x_1,x_2, ..., x_n)
   \end{align*}
   where the $i^{th}$ component of $\boldsymbol{\sigma}(x_1,x_2, ..., x_n)$ is 
   \begin{equation}
       [\boldsymbol{\sigma}(x_1,x_2, ..., x_n)]_i=\frac{e^{x_i}}{\sum_je^{x_j}}
   \end{equation}

   As the output is normalised, it can be interpreted as the probability that the given \ac{cwt} map belongs to any of the classes we have trained over. Being a multiclass classification problem, we use categorical cross-entropy as our loss function, Adam as the optimiser, and the F1-Score\footnote{$F_1=\mathrm{\frac{2TP}{2TP+FP+FN}}$; TP = True Positives, FP = False Positives, FN = False Negatives} as the evaluation metric since it is unaffected by dataset imbalance \cite{Kingma:2014}. Additionally, to mitigate the risk of overfitting, we train with an early-stopping regularisation and use the model displaying the maximum validation F1-Score.

\subsection{Revised ranking statistics and significance} \label{ranking}

We evaluate the trained model on a given trigger, either \ac{gw} transient or noise transient. These CWT maps are constructed separately for the Hanford and Livingston detectors and are subsequently fed into the network. For each of these maps, the model outputs the probability of the trigger corresponding to the signal class, which we denote by $P_{H1}$ and $P_{L1}$ for the Hanford and Livingston detectors, respectively. Motivated by \citep{Sharma:2023nuv}, we use these probabilities to construct the CNN statistic as

   \begin{equation}
     R_\mathrm{CNN}=\frac{1}{2}\ln\left[\left(\frac{P_{\mathrm{H1}}}{1-P_{\mathrm{L1}}}\right)^2+\left(\frac{P_{\mathrm{L1}}}{1-P_{\mathrm{H1}}}\right)^2\right]
    \label{eq:quadsum}
   \end{equation}

 The functional form of this ranking statistic ensures that signals, typically assigned $P_{H1}$ and $P_{L1}$ values close to 1, yield high values of $ R_\mathrm{CNN}$. This occurs because the $1-P_{H1/L1}$ term in the denominator amplifies the effect of high signal probabilities. Conversely, noise triggers, with $P_{H1}$ and $P_{L1}$ values close to 0, result in low values for $R_\mathrm{CNN}$. The resulting separation between signals and noise, driven by $R_\mathrm{CNN}$, plays a key role in enhancing the sensitivity of the search. 

  We thus modify the base search ranking statistic as 
   \begin{equation} 
   R =\begin{cases} 
      R_\mathrm{base}+R_\mathrm{CNN} &  R_\mathrm{base}>\alpha\ \&\  R_\mathrm{CNN}>\beta\\
      R_\mathrm{base} & \mathrm{otherwise}
   \end{cases} \label{eq:ranking_stat}
   \end{equation} 
    The $\alpha$ and $\beta$ parameters are tunable. We impose a threshold on $R_\mathrm{base}$ since certain searches have too large a background for which CWT map generation is infeasible due to the high computational cost associated with it. The threshold on $R_\mathrm{CNN}$, on the other hand, is to prevent signals with high $R_\mathrm{base}$ and low $R_\mathrm{CNN}$ from dropping in significance due to a misclassification from our model. The exact values are tuned based on the specific search.

 Using this ranking statistic, it is possible to assign a significance to coincident triggers by estimating the likelihood that such a trigger could arise from noise. Specifically, the significance is quantified by calculating the rate (also known as the \ac{far}) at which coincident noise triggers would be expected to produce a ranking statistic equal to or greater than that of the observed trigger. One such method to calculate the \ac{far} is the method of time slides\citep{Usman:2015kfa, Chu:2020pjv}, where the data of one detector is shifted relative to another to generate fake trigger coincidences, referred to as background triggers. We apply our reweighing to the background triggers as well to estimate the False Alarm rate of a foreground trigger with ranking statistic value $R^*$ as
    \begin{equation} \label{eq:far}
      \mathrm{FAR}=\frac{n_b(R^*)+1}{T_b}
    \end{equation}
    Here $n_b(R^*)$ is the number of background triggers exceeding the ranking statistic $R^*$, and $T_b$ is the total background time generated by time-sliding.

\section{Demonstration with PyCBC based IMBH searches} \label{application}
\subsection{IMBH playground data} \label{playground}

 In the following sections, we focus our attention on a specific class of \ac{gw} burst signals from coalescing \ac{imbh} binaries. \acp{imbh} are \acp{bh} with masses between \(100 \, M_\odot\) and \(10^5 \, M_\odot\) and are hypothesised to fill the observational gap between stellar mass \acp{bh} and supermassive \acp{bh} \citep{Inayoshi:2019fun}. At low redshifts, these \acp{imbh} can form via runaway collisions of seed \acp{bh} in dense stellar clusters (\citep{Greene:2019vlv} and references within). Initial low mass \acp{bh} will accrue post-merger spin via these collisions, leading to subsequent mergers having high spin. If these IMBH seeds form in nuclear star clusters with high escape velocity, there could be multiple hierarchical mergers before the remnant is kicked outside the cluster \citep{Mapelli:2021syv}. Such binaries formed in dynamical scenarios exhibit a greater likelihood of mass asymmetry, isotropic spins and orbital eccentricity as compared to binaries undergoing isolated evolution. 
 These signatures of mass asymmetry and misaligned spins are imprinted onto the observed \ac{gw} waveform in the form of modulations due to additional physical effects such as higher order harmonics, spin-orbit precession and orbital eccentricity. Searching for such systems is challenging due to the complex morphology in the waveforms brought on by the additional physics, as well as the lack of available waveforms which encapsulate all the physical effects in the wide parameter space. Despite this, tuned searches for IMBH binaries in the quasispherical regime exist that capture the impact of higher-order modes \citep{Chandra:2022ixv, Wadekar:2023kym} and spin-orbit precession \citep{Schmidt:2024jbp}.
 

Here, we test the CNN reweighing on two matched-filter based PyCBC searches optimised for quasi-circular IMBH binaries, namely PyCBC-IMBH and PyCBC-HM, that we detail in the following sections \cite{Chandra:2021xvs}, \cite{Chandra:2022ixv}
We train our \ac{ml} algorithm for systems with detector frame total mass between $100 M_\odot$ to $600M_\odot$. Signals exceeding this upper limit do not even complete one cycle within advanced LIGO, making it infeasible to separate them from glitches. Similarly, we also restrict ourselves to mass ratios between $(1,10)$ and aligned spins from $(-0.9, 0.9)$. We plan to explore the impact of misaligned spins in a future work. In the following subsections, we demonstrate the effectiveness of our model using these IMBH searches. 
 
 

\subsection{PyCBC-IMBH} \label{pycbc-imbh}

\begin{figure}
       \includegraphics[width=0.9\columnwidth]{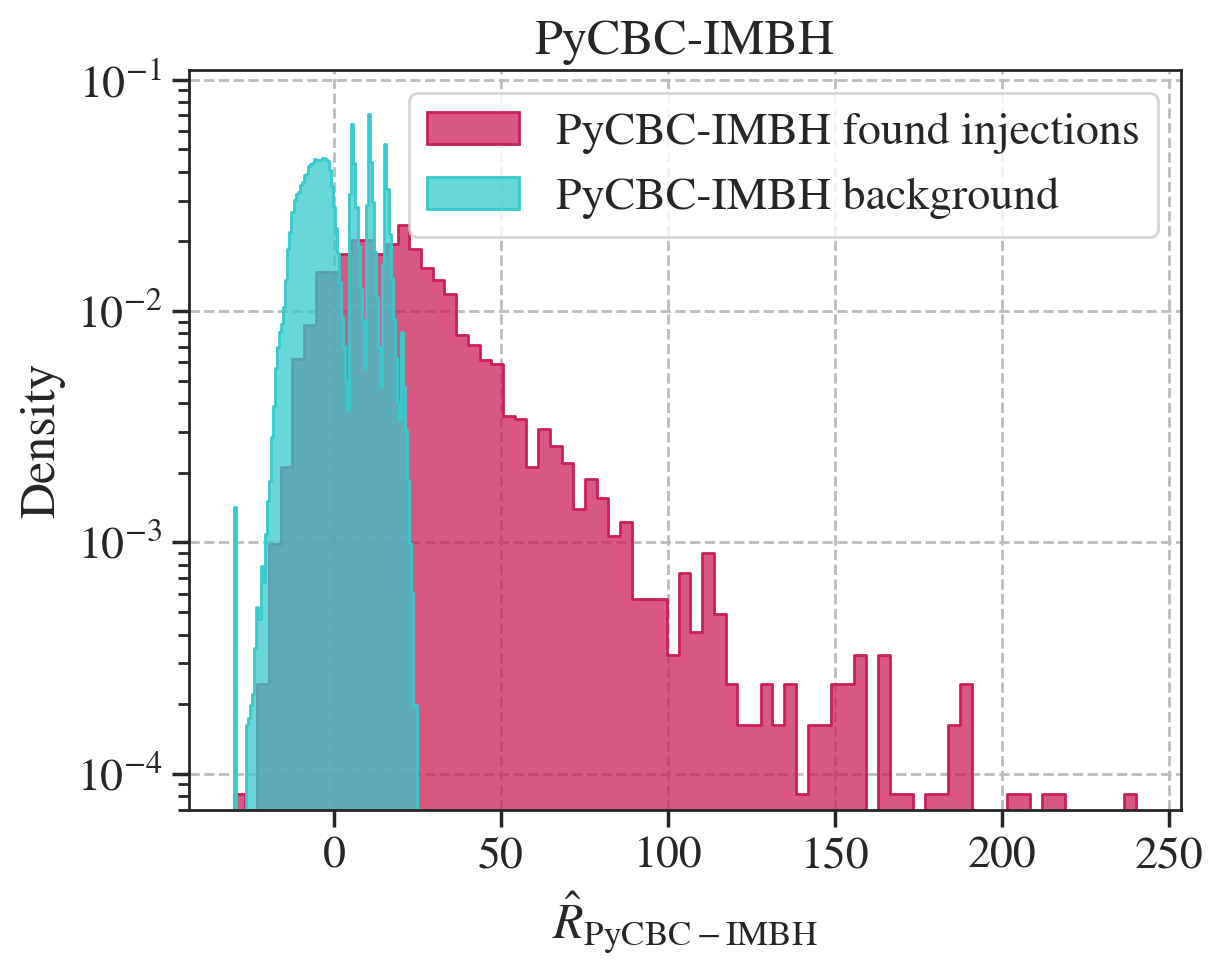}
       \caption{A histogram of the PyCBC-IMBH background triggers overlaid with the injections found by PyCBC-IMBH}
       \label{fig:pycbc_imbh_bkg} 
   \end{figure}

PyCBC-IMBH is a matched filter search tuned for IMBH binary systems utilising a quasicircular quadrupolar template bank. As discussed in Section \ref{motives}, as high mass searches are adversely affected by short duration glitches, PyCBC-IMBH uses strict vetoes to reduce the significance of instrumental transients. Additionally, the template bank mitigates the impact of the {\it look elsewhere} effect by constraining the parameter space, which enhances the sensitivity to IMBH systems, achieving up to three times better performance compared to the broad template bank search used in PyCBC \citep{Chandra:2022ixv}. Figure \ref{fig:pycbc_imbh_bkg} presents the distribution of background triggers, with the recovered injections from the PyCBC-IMBH search overlaid. The background distribution exhibits a sharp cutoff, beyond which no background triggers are observed. This lack of a high-significance tail implies that any trigger exceeding this threshold is assigned the minimum possible false alarm rate (FAR) of $1/T_b$ $\mathrm{yr}^{-1}$. The substantial fraction of injections lying above this cutoff demonstrates the search’s ability to effectively separate signal from noise.

To construct our training dataset, we visually examine CWT maps of background triggers to determine the prominent glitch classes. Our training dataset thus consists of blips, low-frequency blips and tomtes. These, in addition to the Gaussian noise and signal classes, form our training dataset. We ensure that our dataset is balanced to avoid any biases during training. Due to this restriction, we are limited by the number of confident glitches obtained from Gravity Spy. Thus, each class contains about 9000 images. To ensure optimal coverage of the parameter space for the signal class, we construct a stochastic template bank using the IMRPhenomXAS \citep{Pratten:2020fqn} waveform model within the ranges given by Table \ref{tab:inj_training_dataset}.

\begin{figure}
       \includegraphics[width=0.95\columnwidth]{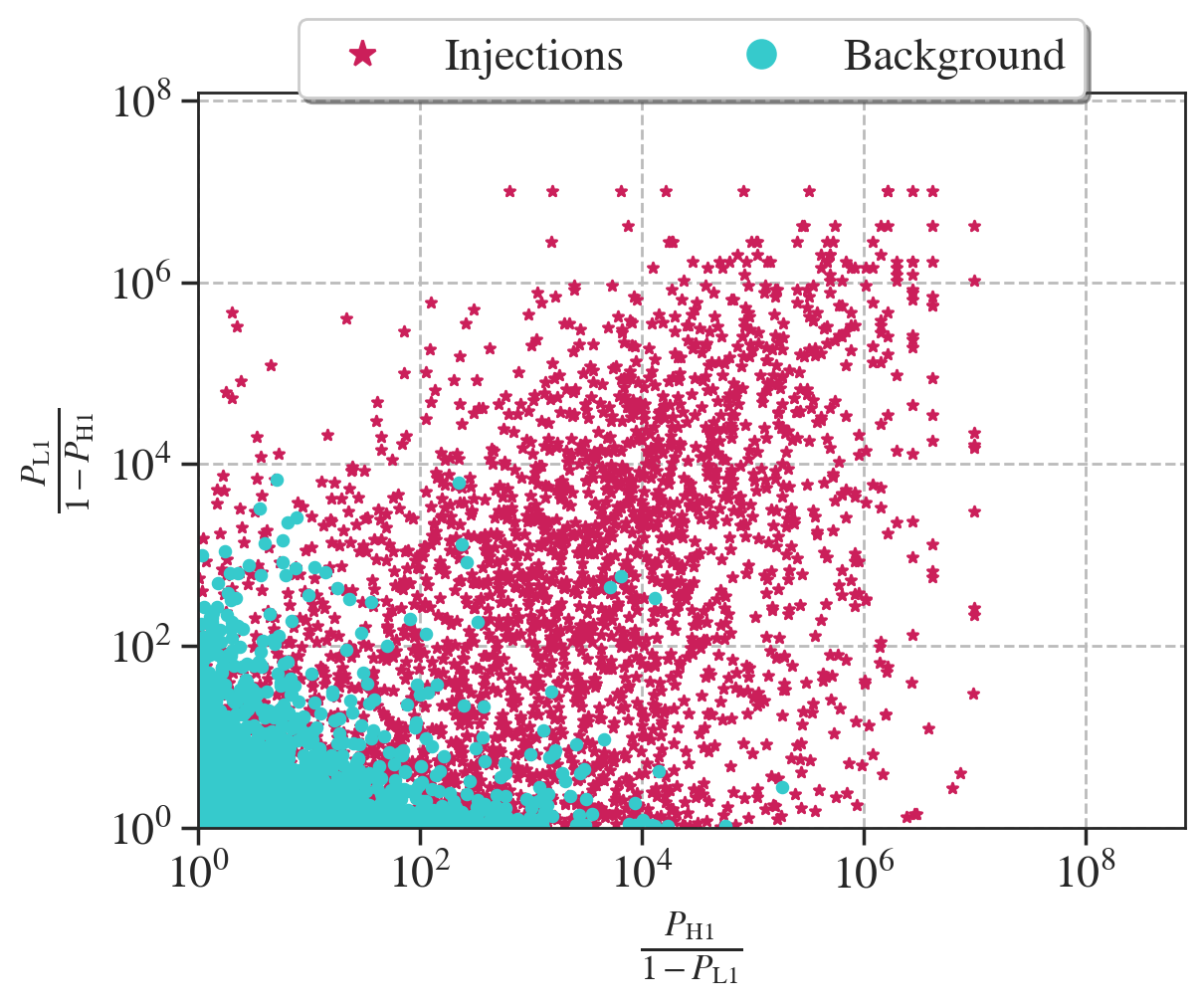}
       \caption{Scatter plot of the background triggers and injections indicating the suppression of noise triggers relative to the injected signals.} 
       \label{fig:inj_bkg_scatter} 
   \end{figure}

While constructing the ranking statistic, we find that the best signal recovery is obtained when no thresholds are imposed on the PyCBC-IMBH ranking statistic and $R_\mathrm{CNN}$. Thus Eq \eqref{eq:ranking_stat} becomes
\begin{equation}
    R_\mathrm{tot}={R}_\mathrm{PyCBC-IMBH}+{R}_\mathrm{CNN}
\end{equation}
To demonstrate the ability of Eq \eqref{eq:quadsum} to separate noise triggers and signals, we apply our model to a single 30 day chunk of O3 data that was analysed by PyCBC-IMBH. About 25,000 injections were made in this chunk with SEOBNRv4\_ROM~\citep{Bohe:2016gbl}, distributed uniformly in the detector frame total mass between $[100M_\odot, 600M_\odot]$ and mass fraction\footnote{$m_1/(m_1+m_2)$} with mass ratio constrained to be between 1 and 10. Additionally, the aligned spins were chosen between $[-0.998, 0.998]$. This choice of intrinsic parameters ensured a comprehensive coverage of our target parameter space, avoiding underrepresentation in any region. Additionally, using a different waveform model for testing ensures that our conclusions are not biased due to the \ac{cnn} learning possible features unique to the waveform model used for training. For the extrinsic parameters, the sky location and inclination angles were drawn from isotropic distributions, while the luminosity distance was sampled uniformly from 0.5 Gpc to 11 Gpc. A uniformly sampled distance ensures that the signals are uniformly distributed in inverse SNR, thus reducing the number of sub-threshold signals as compared to using a uniform-in-volume distribution. The upper limit of the luminosity distance is the horizon distance of our training dataset signals with a single detector SNR threshold of 4, ensuring that we don't generate any signals that are below the detection limit. Figure \ref{fig:inj_bkg_scatter} presents a scatter plot of the PyCBC-IMBH recovered injections and background triggers. The axes of the plot are the components of Eq~\eqref{eq:quadsum}. The background triggers cluster in the lower-left region of the plot, corresponding to low values of $R_\mathrm{CNN}$, whereas the injections are more broadly distributed toward the upper-right, indicating a clear separation between signal and noise for this particular class of the signal and noise transients at the output of the PyCBC-IMBH. This separation provides an additional, independent approach to address the signal-noise transient classification problem. We present quantitative results in Sec. \ref{results}, where we compute the improvement in search sensitivity achieved by our approach.

   \begin{table}[h!]
    \begin{tabular}{|c|c|c|}
        \hline 
        Parameter & PyCBC-IMBH & PyCBC-HM \\
        \hline \hline 
        $M_T(1+z)$ & {$[100,600] M_{\odot}$} & {$[100,600] M_{\odot}$} \\
        $q$ & {$[1,10]$} & {$[1,10]$} \\
        $\chi_z$ & {$[-0.9,0.9]$} & {$[-0.9,0.9]$} \\
        $\iota$ & {$\left[0^{\circ}, 180^{\circ}\right]$} & {$\left[60^{\circ}, 120^{\circ}\right]$} \\
        $\varphi$ & - & {$[0,2 \pi]$} \\
        $\rho_{\text {opt }}$ & {$[5,40]$} & {$[5,40]$} \\
        \hline
    \end{tabular}
    \caption{Details regarding the training dataset.}
    \label{tab:inj_training_dataset}
    \end{table}
\subsection{PyCBC-HM}

Asymmetric, edge-on \ac{imbh} binary systems carry significant signal energy in the higher-order modes. If these modes are not taken into account, there are high chances of missing such binaries as they have immense potential to probe \ac{bh} formation channels.
To combat this PyCBC-HM was designed to improve the detection algorithm for such systems by incorporating \acp{hm} into the quasi-circular template bank. This extension has shown a threefold improvement in sensitivity as compared to the PyCBC-IMBH search; clearly demonstrating that PyCBC-HM is particularly beneficial for detecting asymmetric, edge-on systems, compared to their mass-symmetric, face-on counterparts \citep{Chandra:2022ixv}. PyCBC-HM targets a more restricted parameter space of edge-on systems with inclination angles between $75^\circ$ and $105^\circ$ as they have a greater contribution from \acp{hm}. Moreover, to limit short-duration templates from picking up glitches, a minimum template duration cut is applied, resulting in the bank having a maximum total mass of $500M_\odot$. 

\begin{figure}
       \includegraphics[width=0.95\columnwidth]{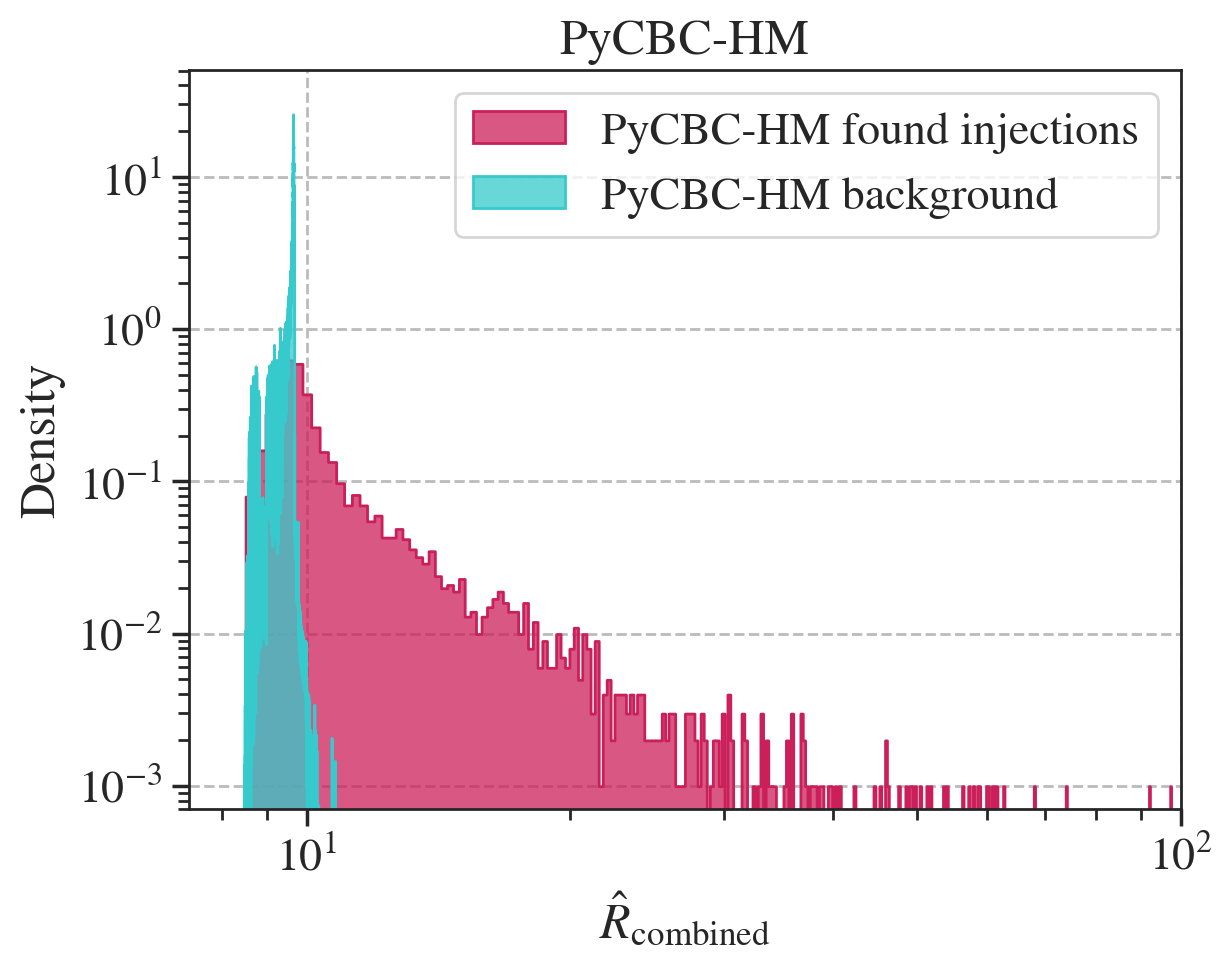}
       \caption{A histogram of the PyCBC-HM background triggers overlaid with a distribution of the injections found by PyCBC-HM}
       \label{fig:pycbc_hm_bkg} 
   \end{figure}

The inclusion of higher-order modes exposes the search to more complex glitches, such as whistles, koi fish, and repeating blips, which possess intricate morphological features. Consequently, the background becomes substantially larger, encompassing approximately 50 million triggers as compared to the 200,000 triggers in PyCBC-IMBH. Despite the increased background size, the search is capable of segregating these noise triggers from signals as is evident from Figure \ref{fig:pycbc_hm_bkg}. The suppression of the tail in the background distribution is achieved by employing several strategies. For instance, PyCBC-HM imposes a maximum threshold on the reduced chi-squared, where triggers exceeding a value of 10 are discarded. Additionally, a modified template-dependent reweighing scheme is employed in the multi-detector ranking statistic, which downranks triggers from templates more susceptible to glitch contamination (see Appendix \ref{appA} for further details).

Since the PyCBC-HM search is capable of detecting additional glitches with complex morphology, it is crucial to include these glitches in the training set. Consequently, our training dataset comprises of a diverse range of noise and glitch types, such as blips, low-frequency blips, tomtes, koi fish, whistles, as well as both signals in Gaussian noise and O3 coloured Gaussian noise obtained from the H1 and L1 detectors Given that PyCBC-HM incorporates higher-order modes, we find that constructing a template bank with a minimal match of 0.97—despite focusing solely on edge-on systems—results in approximately 40,000 templates. Although lowering the minimal match criterion would reduce the number of templates, this approach poses a challenge as templates become increasingly sparse at higher total masses. Consequently, a reduced match threshold leads to an underrepresentation of signals in this region, which negatively impacts training for higher-mass systems. To address this, the signals are sampled uniformly across the parameter space specified in Table \ref{tab:inj_training_dataset}, ensuring adequate representation throughout the entire mass spectrum.

In contrast to the PyCBC-IMBH search, we apply the reweighing procedure to only a subset of triggers in the PyCBC-HM search. This decision is driven by the significantly larger number of background triggers in PyCBC-HM (approximately $10^7$), which makes generating CWT maps for all of them computationally prohibitive. To manage this cost, we set $\alpha=9.8$, and only generate CWT maps for triggers with a PyCBC-HM ranking statistic above this value. The choice of $\alpha$ was informed by the number of background triggers exceeding it across multiple O3a PyCBC-HM analysis chunks. We aimed to limit the total number of CWT map generations to a manageable level, and found that $\alpha=9.8$ yields approximately 200,000 background triggers, an amount that is computationally feasible and broadly comparable to the full background of PyCBC-IMBH. Moreover, we choose $\beta=0$ since certain signals were assigned very low values of $R_\mathrm{CNN}$, leading their IFAR to drop considerably. With these values Eq \eqref{eq:ranking_stat} becomes

   \begin{equation} 
   R_\mathrm{tot}=\begin{cases} 
      R_\mathrm{PyCBC-HM}+R_\mathrm{CNN} &  R_\mathrm{base}>9.8\ \&\  R_\mathrm{CNN}>0\\
      R_\mathrm{PyCBC-HM} & \mathrm{otherwise}
   \end{cases}
   \end{equation}
We apply these to the PyCBC-HM search triggers to estimate the improvement in search sensitivity and present the results in the following section.

    \section{Results} \label{results}
  Searches are typically evaluated by computing the sensitive volume time product (VT), which is a measure of the average spacetime volume the search is sensitive to at a given significance threshold. It is given by the following expression \citep{Usman:2015kfa, LIGOScientific:2016ebi}

  \begin{equation}
      \mathrm{VT}(\mathcal{F})=T\int \epsilon(\mathcal{F};{\bf x, \Lambda})p({\bf x, \Lambda})\mathrm{d}{\bf x}\mathrm{d}{\bf \Lambda}
  \end{equation}
  where ${\bf\Lambda}$ are the intrinsic parameters of the signal, $\epsilon$ is the detection efficiency at a FAR of 
 $\mathcal{F}$, ${\bf x}$ is the source location, $p({\bf x, \Lambda})$ is the distribution of the signal sources and $T$ is the total search time. In the following subsections, we will compare the improvement in sensitive VT for the PyCBC-IMBH and PyCBC-HM searches.

    \subsubsection{PyCBC-IMBH}
\begin{figure*}[hbt!]
       \includegraphics[scale=0.15]{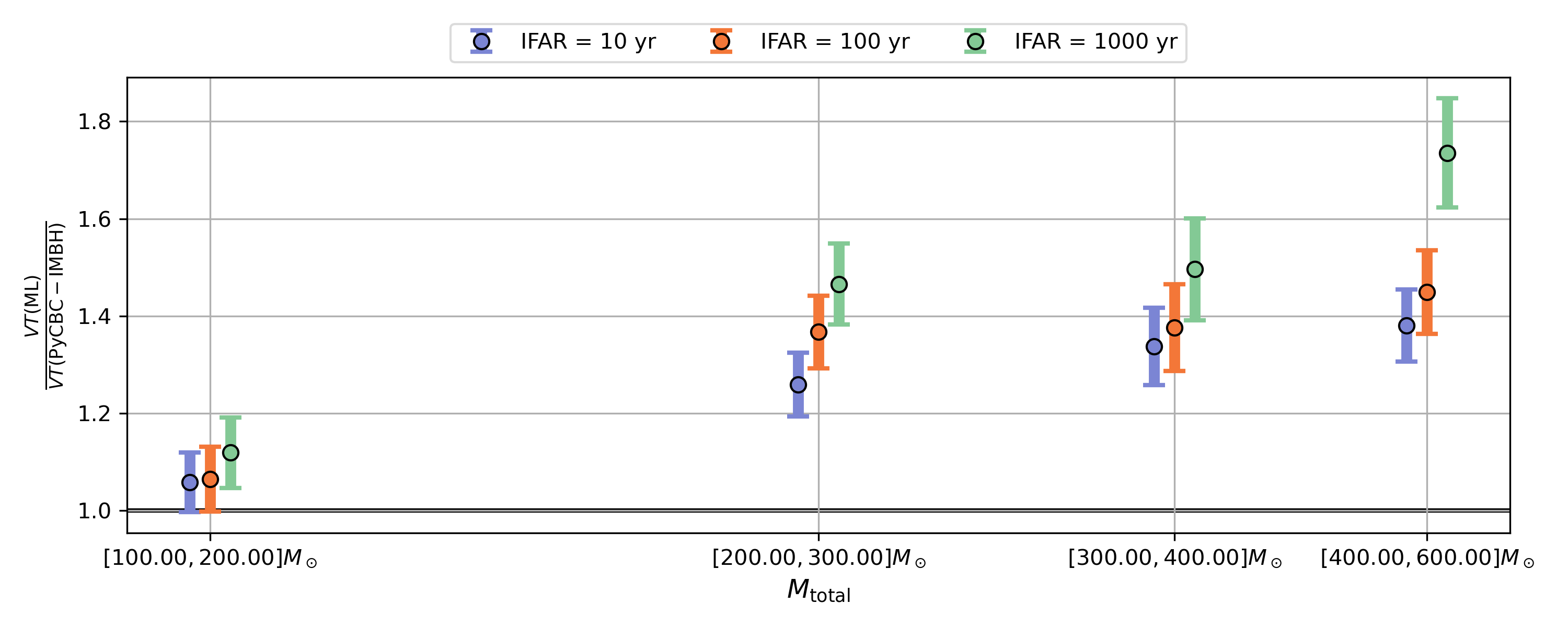}     \caption{The sensitive volume time product ratio between PyCBC-IMBH and its ML assisted version}
       \label{fig:vt_ratio_imbh}
   \end{figure*}

     The PyCBC-IMBH sensitivity analysis was performed over the entirety of O3a data in separate 30 day chunks. To demonstrate the efficacy of our algorithm, we apply our model to the first four of these chunks, ranging from the  1st of April to the 24th of August 2019. For each of these chunks, we distribute the signal parameter as detailed in sec \ref{pycbc-imbh}. 
     Fig \ref{fig:vt_ratio_imbh} plots the ratio of the sensitive volume times binned by the total mass of the injections. We note that at an IFAR of 100yr the VT ratio improvement ranges from 15\% for the $[100M_\odot,200M_\odot)$ mass bin to about 70\% for the $[400M_\odot, 600M_\odot)$ mass bin. 
     
     We also assess the significance of real events from the first four O3a chunks using the new ranking statistics. The event significance is computed with the reweighed background from its respective chunk. As PyCBC-IMBH was originally run for the Hanford-Livingston-Virgo network rather than the two detector network (Hanford and Livingston) our model expects, we report the PyCBC-IMBH event significances with respect to the Hanford-Livingston background. Tab \ref{tab:openbox_pycbc_imbh} lists the events along with their IFAR values as recovered by PyCBC-IMBH and its \ac{ml} assisted version. Most of the real events have improved IFAR and we recover the first confident IMBH binary event GW190521 with IFAR of $> 42000$ years, almost 20 times higher than the optimised PyCBC-IMBH search. A few events drop in significance values which we attribute to their low detector frame total mass ($<100M_\odot$), placing them outside our training parameter space. However GW190706\_222641 also has a detector frame total mass below $100M_\odot$ and its ranking statistic does decrease after reweighing. Nevertheless, it remains above the background distribution and thus the lower limit on its IFAR remains intact.

   \begin{table}[h!]
    \begin{tblr}{colspec = {|c|c|c|}}
        \hline 
        Event & IFAR PyCBC-IMBH & IFAR ML \\
        \hline \hline 
        GW190706\_222641 & $>52836.6$ & $>52836.6$ \\
        GW190521 & 2140.6 & $>42813.2$ \\
        GW190521\_074359 & $>42813.2$ & $>42813.2$  \\
        GW190519\_153544 & $>42813.2$ & $>42813.2$ \\
        GW190727\_060333 & $>66722.0$ & 33361.0 \\
        GW190602\_175927 & 2518.4 & 21406.6 \\
        GW190503\_185404 & 3878.4 & 633.2 \\
        GW190421\_213856 & 172.4 & 254.3 \\
        GW190517\_055101 & 4281.3 & 236.5 \\
        GW190408\_181802 & 585.4 & 56.3 \\
        GW190513\_205428 & 37.5 & 53.1 \\
        \hline
    \end{tblr}
    \caption{Applying our algorithm to real O3a events and comparing their IFARs in per year with PyCBC-IMBH. Some of the events are reported with a lower limit on IFAR due to the finite background statistics.}
    \label{tab:openbox_pycbc_imbh}
    \end{table}

   \subsubsection{PyCBC-HM}
   The PyCBC-HM injections' intrinsic parameters were distributed identically to those of PyCBC-IMBH with the exception of the luminosity distance, which was uniformly distributed in comoving volume from $0.26\ \mathrm{Gpc}^3$ to $40\ \mathrm{Gpc}^3$ and the inclination angle, which was constrained to lie between $75^\circ$ and $105^\circ$. As in the previous analysis, we focus on the first four chunks of O3a.

\begin{figure*}[hbt!]     
       \includegraphics[scale=0.15]{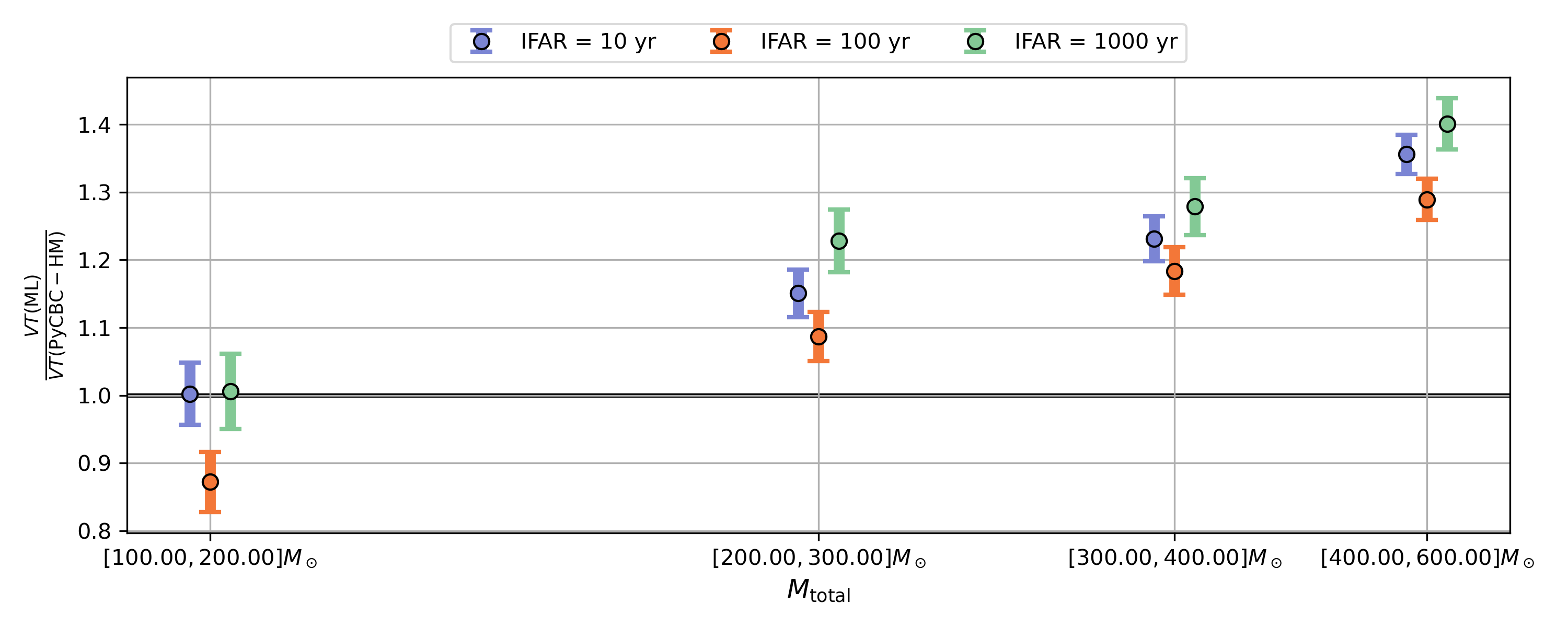}
       \caption{The sensitive volume time product ratio between PyCBC-HM and its ML assisted version}
       \label{fig:vt_ratio_hm}
   \end{figure*}

   Fig \ref{fig:vt_ratio_hm} shows the VT ratio comparing PyCBC-HM with its ML assisted version. While there is a noticeable drop in performance for the $[100M_\odot, 200M_\odot)$ mass bin, a general trend of increasing performance with higher mass bins is observed. At an IFAR of 100 years, we observe a maximum improvement in performance of about 30\% in the $[400M_\odot, 600M_\odot)$ mass bin. The performance drop in the lower mass bins, especially at an IFAR of 100 yr suggests that our model faces challenges in dealing with longer duration signals. This is partly because at low masses the \ac{cnn} model is outperformed by matched filtering. Another contributing factor was the use of a stochastic template placement for the training dataset. Since the match for low masses drops faster compared to high masses, our choice of a uniform mass distribution did not adequately cover the low mass region. As a result, most low mass signals drop below the 100 yr IFAR mark, hovering above an IFAR of 10 yr, leading to the observed decrease in VT. As a potential solution for mitigating this drop one could explore a better informed training dataset where the lower mass bin is better represented.


   As in the previous analysis, we construct CWT maps of triggers found by PyCBC-HM. Table \ref{tab:openbox_pycbc_hm} presents the IFARs of real gravitational wave (GW) events recovered by the PyCBC-HM search, along with their IFARs after being processed by our model. We observe an improvement in IFARs for the majority of events, with the exceptions of GW190521\_074359, GW190408\_181802, and GW190519\_153544. The first two events have a detector-frame total mass below $100M_\odot$, which is outside the scope of the training dataset. In contrast, GW190519\_153544 has a detector-frame total mass of $153M_\odot$, which falls within the CNN dataset. The reduced IFAR is attributed to the signal being misclassified as a blip in the L1 detector. This is owed to the morphological similarity between blips and \ac{imbh} signals, and in such cases, it can reduce the significance of true signals. This issue is not unique to our method, as other excess power unmodelled approaches also capture blips when looking for massive \ac{bbh} signals. Other independent approaches need to be explored in these cases.

      \begin{table}[h!]
    \begin{tabular}{|c|c|c|}
        \hline 
        Event & IFAR PyCBC-HM & IFAR ML \\
        \hline \hline 
        GW190521\_074359 & $598.2$ & $125.2$  \\
        GW190513\_205428 & $0.23$ & $14.1$ \\
        GW190706\_222641 & $0.46$ & 7.18\\
        GW190503\_185404 & 0.31 & 6.59\\
        GW190408\_181802 & 10.49 & 5.19 \\
        GW190519\_153544 & 18.82 & 2.06 \\
        GW190727\_060333 & 0.58 & 1.30 \\
        \hline
    \end{tabular}
    \caption{Applying our algorithm to real O3a events and comparing their IFARs in per year to PyCBC-HM}
    \label{tab:openbox_pycbc_hm}
    \end{table}
       

\section{Conclusions}
In this study, we propose using a \ac{cnn} to augment existing searches by modifying their ranking statistic. In particular, we exploit the ability of CNNs to distinguish between \ac{cwt} maps of glitches and signals. We use the network's output to downrank coincident noise triggers using a carefully constructed statistic as outlined in \citep{Sharma:2023nuv}. 
We demonstrate the efficacy of this approach in a generic manner on two PyCBC based searches targeting IMBH binary systems. Despite these searches already being highly optimised, we show that our approach further improves search sensitivity across all mass bins for PyCBC-IMBH and at higher masses for PyCBC-HM.

When reweighing real O3a gravitational wave (GW) events, our model consistently recovers signals with higher significance. Exceptions are observed primarily when event parameters lie outside the training domain, or, in the case of PyCBC-HM, when a strong degeneracy exists between the signal and blip-like noise transients. Although we demonstrated our method on two matched-filter based searches, the approach is not inherently limited to these techniques and can be readily applied to unmodelled searches as well. This work presents a proof of concept for augmenting existing GW search pipelines using \acp{cnn}, demonstrating that they can be seamlessly incorporated as an independent signal noise discriminator for a chosen \ac{gw} transient search. 

    \section*{Acknowledgements}

The authors thank Rahul Dhurkunde for his comments and valuable suggestions, which greatly improved this paper. JF acknowledges support from the Prime Minister's Research Fund and SPARC/2019-2020/P2926/SL, Government of India for the travel support to Pennsylvania State University during which part of the project was carried out. AP acknowledges the support from SPARC MoE grant SPARC/2019-2020/P2926/SL, Government of India. KC acknowledges the generous support provided through NSF grant numbers PHY-2207638, AST-2307147, PHY-2308886, and PHY-2309064. The authors are grateful for computational resources provided by the LIGO Laboratory (LLO cluster) and supported by the National Science Foundation Grants PHY-0757058 and PHY-0823459. 
This research has made use of data, software and/or web tools obtained from the Gravitational Wave Open Science Center (https://www.gw-openscience.org),
a service of LIGO Laboratory, the LIGO Scientific Collaboration, the Virgo Collaboration, and KAGRA. This material is based upon work supported by NSF’s LIGO Laboratory, which is a major facility fully funded by the National Science Foundation. LIGO Laboratory and Advanced LIGO are funded by the United States National Science Foundation (NSF) as well as the Science and Technology Facilities Council (STFC) of the United Kingdom, the Max-Planck-Society (MPS), and the State of Niedersachsen/Germany for support of the construction of Advanced LIGO and construction and operation of the GEO600 detector. The Australian Research Council provided additional support for Advanced LIGO. Virgo is funded through the European Gravitational Observatory (EGO), by the French Centre National de Recherche Scientifique (CNRS), the Italian Instituto Nazionale di Fisica Nucleare (INFN) and the Dutch Nikhef, with contributions by institutions from Belgium, Germany, Greece, Hungary, Ireland, Japan, Monaco, Poland, Portugal, Spain. KAGRA is supported by the Ministry of Education, Culture, Sports, Science and Technology (MEXT), Japan Society for the Promotion of Science (JSPS) in Japan; National Research Foundation (NRF) and the Ministry of Science and ICT (MSIT) in Korea; Academia Sinica (AS) and National Science and Technology Council (NSTC) in Taiwan. 
%

\appendix
\section{PyCBC based IMBH searches}
\label{appA}
    PyCBC searches use matched filtering to extract a signal $\boldsymbol{s}$ from the detector data $\boldsymbol{d}$. They construct a collection of quasicircular quadrupolar waveforms $\boldsymbol{h}$ referred to as a template bank; parametrised by the component masses and the aligned spin components along the orbital angular momentum. However, if templates include higher order harmonics such as the PyCBC-HM search, then the template bank incorporates the additional parameters of the binary inclination and azimuth. Given $\boldsymbol{h}$, the most generic matched-filter \ac{snr} statistic capable of identifying any $\boldsymbol{s}$ from $\boldsymbol{d}$ is \citep{Harry:2017weg}:  
\begin{widetext}  
    \begin{equation}  \label{eq:generic-snr}
        \rho^2 := \frac{\left(\boldsymbol{d} \mid \hat{\boldsymbol{h}}_{+}\right)^2 + \left(\boldsymbol{d} \mid \hat{\boldsymbol{h}}_{\times}\right)^2 - 2\left(\boldsymbol{d} \mid \hat{\boldsymbol{h}}_{+}\right)\left(\boldsymbol{d} \mid \hat{\boldsymbol{h}}_{\times}\right)\left(\hat{\boldsymbol{h}}_{+} \mid \hat{\boldsymbol{h}}_{\times}\right)}{1-\left(\hat{\boldsymbol{h}}_{+} \mid \hat{\boldsymbol{h}}_{\times}\right)^2}.
    \end{equation}  
\end{widetext}  

Here, $\hat{\boldsymbol{h}}_{+/\times}$ are the unit-normalised template polarizations in the frequency domain and \(\left(\boldsymbol{d} \mid \hat{\boldsymbol{h}}_{+/\times}\right)\) is given by:
\begin{equation}
    \left(\boldsymbol{d} \mid {\boldsymbol{h}}_{+/\times}\right)=4\Re \left(
    \sum_{k}\frac{\boldsymbol{\tilde d}(f_k)^*
    \boldsymbol{\tilde h_{+/\times}}(f_k)}{S_n(f_k)}\Delta f\right)
\end{equation}

Here $\Re$ stands for the real part of a complex number, $S_n(f_k)$ is the noise \ac{psd} at frequency $f_k$, $\Delta f$ is the frequency step and the tilde denotes the Fourier transform of the corresponding time domain data. By construction, this statistic \textit{effectively} maximises over the 3D sky-location parameters of the source and the polarisation angle. For searches using quasi-circular quadrupole templates, such as PyCBC-IMBH, Eq.~\eqref{eq:generic-snr} is 
\begin{equation}\label{eq:snr}
    \rho^2 = \left(\boldsymbol{d} \mid \hat{\boldsymbol{h}}_{+}\right)^2 + \left(\boldsymbol{d} \mid \hat{\boldsymbol{h}}_{\times}\right)^2 
\end{equation}
as $\hat{\boldsymbol{h}}_{+} := i\hat{\boldsymbol{h}}_{\times}$. This statistic, although restrictive, is computationally more efficient than Eq.~\eqref{eq:generic-snr} as it requires only one noise-weighted inner product computation. To deal with instrumental transients PyCBC searches evaulate triggers based on multiple criteria, including whether they (a) are a significant outlier in the calibrated, whitened data stream, (b) exhibit a morphology consistent with the best-matched template waveform, (c) show no excess power above the maximum frequency of the best-matched template, (d) lack any significant excess power summed across frequency bands on characteristic timescales, and (e) demonstrate coincidence across multiple detectors. These conditions appear as a reweighing factor that suppresses the SNR statistic for noise triggers. 

Triggers are identified as multi-detector candidates when at least two detectors register a single-detector trigger with the same best-matched template within a specified time window. This time window exceeds the light travel time between the detectors by a few milliseconds ~\citep{Usman:2015kfa} to accommodate measurement uncertainty. Together, the template and temporal coincidence criteria effectively exclude most glitches as they are local artefacts. The O3 PyCBC-IMBH search assigns these multi-detector candidates the following ranking statistic \cite{Davies:2020tsx}:
      \begin{equation}
          \Lambda_s=-\ln A_{\{a\}}-\sum_a \ln r_{a,i}(\breve\rho_a)+\ln\frac{p(\vec\Omega|S)}{p(\vec\Omega|N)}+3\ln\frac{\sigma_{\min}}{\sigma_{\mathrm{ref}}}~. \label{eq:multi-detector}
      \end{equation}
      The first two terms account for the coincident noise rate densities dependent on the allowed time window $A$ and the single detector ranking statistic $r_{a, i}$ in the detector $a$ and template $i$. The third term includes the probability distributions of the extrinsic signal parameters $\vec\Omega$ consisting of the trigger time delay, the relative phase difference, and amplitude ratios under the signal and noise hypotheses across the network. The final term penalises the ranking statistic based on the least sensitive detector relative to a reference sensitivity.

\nocite{*}

\bibliography{apssamp}

\end{document}